\documentclass[conference]{IEEEtran}
\IEEEoverridecommandlockouts
\usepackage[T1]{fontenc}
\usepackage{cite}
\usepackage{amsmath}
\usepackage{graphicx}
\usepackage{booktabs}
\usepackage{xurl}
\usepackage{balance}
\graphicspath{{figures/}}
\title{Can STATCOM and UPS-Based Voltage Support Mitigate Fault-Induced Reductions in AI Data Center Grid Power Draw?}
\author{
\IEEEauthorblockN{
Ziang Liu\textsuperscript{1}, 
Yimeng Liu\textsuperscript{1}, 
Francis Yunhe Hou\textsuperscript{1}, 
\IEEEauthorblockA{
\textsuperscript{1}\textit{Department of Electrical and Computer Engineering,} \textit{The University of Hong Kong, Hong Kong, China} \\
liuziang1119@163.com, liuyimen25@connect.hku.hk, yhhou@eee.hku.hk} }

}
\begin{document}
\maketitle

\begin{abstract}
Artificial intelligence data centers (AIDCs) are rapidly increasing in number and capacity. These large electronic loads are interfaced through power electronic devices and can sharply reduce grid power intake during fault-induced voltage sags. Disturbances exceeding equipment ride-through capabilities may also trigger protective disconnection, posing risks to power system security. 
Static synchronous compensators (STATCOMs) and appropriately controlled AIDC uninterruptible power supply (UPS) rectifiers can provide reactive power support during voltage sags.
This paper compares the individual and combined effectiveness of STATCOM and UPS-based voltage support in mitigating fault-induced reductions in grid power intake through electromagnetic transient simulations of a 1 GW AIDC connected to an IEEE nine-bus system.
Simulation results show that STATCOM support better preserves short-term intake in milder tested faults, whereas UPS coordination maintains rectifier operation under deeper sags by reducing active intake to reserve reactive current capacity. Combined support improves auxiliary load retention and recovery in selected cases. However,  severe tested sags still cause substantial grid-demand reductions even with combined support.
\end{abstract}

\section{Introduction}

The rapid growth of AIDCs is increasing the concentration of large power-electronic loads in transmission networks~\cite{mitra2023,miah2026voltage}. Voltage disturbances can sharply reduce their grid power intake even when batteries maintain internal supply. NERC documented approximately 1500~MW of data center load reduction during the July 2024 Eastern Interconnection disturbance~\cite{nerc2025}. Such events highlight the importance of both fault-time demand reduction and post-fault recovery.

STATCOMs have been used to enhance wind farm fault ride-through, and coordinated control with wind farm converters can improve post-fault damping~\cite{molinas2008lvrt,qiao2009coordination}. Converter-based strategies allocate limited current capacity between active and reactive components, often prioritizing reactive support during voltage sags~\cite{thapa2016lvrt,camacho2015current}. The allocation and restoration of active current also affect post-fault voltage recovery~\cite{stankovic2022recovery}. These studies establish the importance of considering reactive support together with converter current limits and active power recovery.

For data centers, grid-forming battery systems and external battery energy storage have been investigated for voltage sag mitigation and voltage stability enhancement~\cite{azizi2025,miah2026voltage}. Coordination of internal resources has also been proposed to improve low-voltage ride-through~\cite{xie2025lvrt}. Uninterruptible power supply (UPS) supervisory control combines active and reactive current allocation, battery buffering, and controlled recovery~\cite{shamseldein2026}. Recent work further allocates grid-side reactive support according to voltage--duration ride-through margins~\cite{lu2026ridethrough}.

For AIDCs, the benefit of voltage support depends on how converter control and protection shape grid power intake. 
Momentary cessation (MC) is a protective converter response in which current exchange is temporarily suspended during severe voltage disturbances without necessarily disconnecting the equipment from the grid~\cite{nerc2018ibr}. In the UPS considered here, MC blocks the rectifier bridge while the battery and dc link continue supplying the downstream IT load~\cite{pnnl2025}.
MC reset and current walk-in can prolong reduced grid intake beyond the voltage sag. Avoiding MC preserves controlled rectifier operation, although UPS support may itself reduce active intake to reserve reactive-current capacity. Auxiliary-load tripping can independently limit the recovery of total facility demand. Comparing external compensation and UPS coordination therefore requires assessment of their effects throughout the fault and recovery.

This paper compares baseline UPS control, STATCOM support, coordinated UPS control, and combined support using electromagnetic transient simulations of a 1~GW PNNL AIDC model connected to an IEEE nine-bus system in PSCAD/EMTDC. The UPS strategy uses a fixed active power reference during voltage support and allocates the remaining current capacity to reactive support. Variations in fault location, resistance, and duration are used to assess MC avoidance, grid-demand reduction, power recovery, auxiliary-load retention, and battery discharge. Additional deep-sag cases examine the limits of combined support.

\begin{figure*}[!t]
 \centering
 \includegraphics[width=\textwidth]{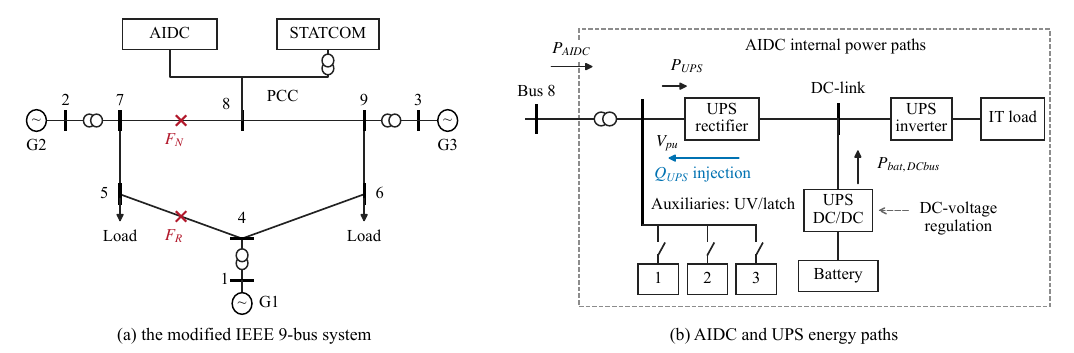}
 \caption{System topology and AIDC power paths. $F_N$ and $F_R$ denote midpoint faults on lines 7--8 and 4--5. 
 }
 \label{fig:system}
 \vspace{-2ex}
\end{figure*}

\section{System Modeling and Voltage Support}
\subsection{Study System and Power Paths}

The 60 Hz IEEE nine-bus system connects a 1 GW AIDC at the Bus 8 point of common coupling (PCC), with a parallel STATCOM through a separate 230/25 kV transformer (Fig.~\ref{fig:system}). The PNNL DC2 model aggregates 1050 UPS units, each using a 1 MVA base at 0.48 kV line-RMS and 1.2 kV dc, with a rectifier, dc link, output inverter, and battery-connected bidirectional dc/dc converter~\cite{pnnl2025}.

Native MC blocks the rectifier bridge when the minimum local phase-RMS voltage, $V_{\mathrm{pu}}$, remains below 0.7 pu for 10 ms. MC resets after local voltage exceeds 0.9 pu for 50 ms. The native current walk-in setting is 4 s; coordinated cases additionally use the soft return below. MC issues no PCC-opening command, while the battery can maintain internal supply. Other rectifier undervoltage, overvoltage, and unbalance trips are disabled. Facility disconnection, ride-through compliance, and system-frequency consequences are outside the study scope.

Three auxiliary branches are supplied directly from the ac distribution system and use 0.75 pu/50 ms undervoltage protection with latched trips and no automatic reclosure. Two three-phase contactors monitor minimum line-to-line RMS voltage and retain 0.15 pu/250 ms unbalance protection; the third branch uses independent single-phase protection.

\begin{figure*}[!t]
 \centering
 \includegraphics[width=\textwidth]{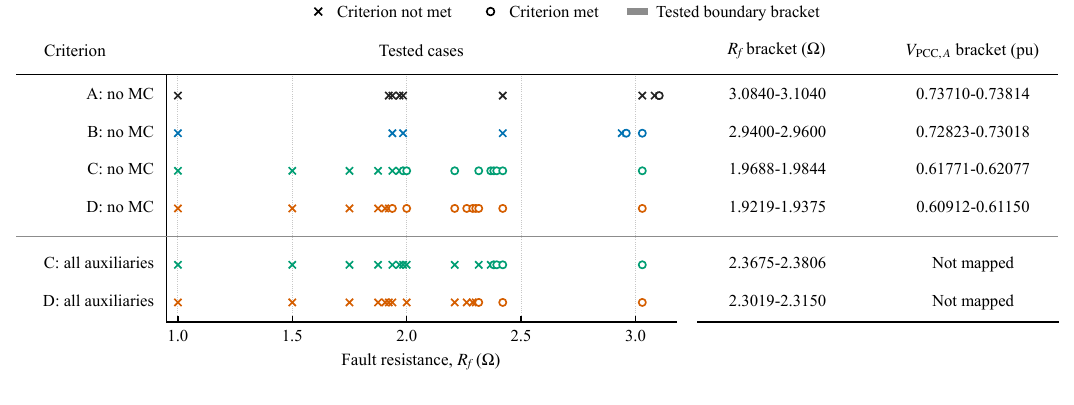}
 \caption{Near-fault MC-avoidance and auxiliary-retention boundaries for 100 ms faults. 
Right-hand values give uncompensated configuration PCC voltage intervals at the MC endpoints. Auxiliary states are reconstructed from protection logic.}
 \label{fig:boundaries}
 \vspace{-2ex}
\end{figure*}

\begin{table*}[!t]
 \caption{Representative Fault Responses of Four Configuration A / B / C / D}
 \label{tab:cases}
 \centering
 \footnotesize
 \begin{tabular}{@{}lcll@{}}
 \toprule
 Fault: location, $R_f$, command & MC duration (ms) & All auxiliaries retained & $E_{\mathrm{bat,+}}$ (MJ)\\
 \midrule
 Near, 3.03 $\Omega$, 100 ms & 93.4 / 0 / 0 / 0 & No / No / Yes / Yes & 933.2 / 7.2 / 139.1 / 136.7\\
 Near, 3.03 $\Omega$, 150 ms & 142.1 / 0 / 0 / 0 & No / No / Yes / Yes & 965.2 / 7.9 / 167.4 / 164.5\\
 Near, 3.03 $\Omega$, 200 ms & 210.1 / 75.7 / 0 / 0 & No / No / Yes / Yes & 1010.5 / 921.6 / 293.5 / 193.7\\
 Near, 2.42 $\Omega$, 100 ms & 135.0 / 135.0 / 0 / 0 & No / No / Yes / Yes & 961.2 / 964.2 / 145.2 / 142.0\\
 Remote, 1.87 $\Omega$, 100 ms & 84.3 / 64.8 / 0 / 0 & No / No / Yes / Yes & 926.3 / 914.9 / 144.2 / 142.4\\
 Near, 1.00 $\Omega$, 100 ms & 231.7 / 279.6 / 232.2 / 282.2 & No / No / No / No & 1028.7 / 1066.3 / 1029.7 / 1068.9\\
 \bottomrule
 \end{tabular}
 \par\smallskip
 \begin{minipage}{\textwidth}\footnotesize
 A: baseline; B: STATCOM; C: UPS coordination; D: combined support. MC and battery discharge use 6.50--11.00 s. ``No'' denotes at least one auxiliary trip, not facility disconnection or an IT outage.
 \end{minipage}
\end{table*}

\subsection{STATCOM and UPS Coordination}

Two voltage-support strategies are considered: external reactive power compensation using a STATCOM and internal reactive-current support through UPS coordination. The STATCOM is represented by an average-value model with AC voltage, DC-link voltage, and inner current control loops. Its rated capacity is 300 Mvar, with a 1.03 pu voltage reference, 0.03 pu droop, and 0.1 pu transformer leakage reactance. The DC-link capacitor is 3000 $\mu$F at 45 kV. The current-reference limit is set to 0.96 pu, while measured-current protection allows 1.0 pu continuously and 1.1 pu for up to 150 ms cumulatively.

The UPS-based voltage support strategy adopted in this study follows the supervisory control concept proposed in~\cite{shamseldein2026}. During a voltage sag, the controller coordinates the active and reactive currents of the UPS rectifier while the battery compensates for the resulting active power deficit through dc-link regulation. A fixed active power reference of $P^*=0.20$~MW per UPS is used, compared with an internal demand of approximately 0.647~MW per UPS. The active-current component is determined first, and the remaining converter current capacity is allocated to capacitive reactive power support. The active power reference remains unchanged during the disturbance, while the battery supplies the power difference required to maintain the internal load.

Let \(v_d\) and \(v_q\) denote the rectifier terminal voltages in the synchronous \(dq\) frame, with \(V=\max(0.05,\sqrt{v_d^2+v_q^2})\). The normalized active-power reference is \(p^{*}=P^{*}/S_b\), where \(P^{*}=0.20\) MW and \(S_b=1\) MVA. The available current magnitude is $I_c=\min\{\max(I_{\lim},0),1\}$.

The controller first reserves active current and allocates the remaining capacity to capacitive voltage support:
\begin{align}
i_a &= \min\!\left(I_c,\frac{p^{*}}{V}\right), \\
i_{r,\max} &= \sqrt{\max\!\left(0,I_c^2-i_a^2\right)}, \\
i_r &=-\min\!\left[i_{r,\max},\,2.5\max(0,1-V_{\mathrm{pu}})\right].
\label{eq:allocation}
\end{align}
Here, negative \(i_r\) denotes capacitive reactive-current injection. The corresponding \(dq\)-axis references are
\begin{equation}
i_d^{*}=\frac{v_di_a-v_qi_r}{V},\qquad
i_q^{*}=\frac{v_qi_a+v_di_r}{V},
\end{equation}
and are tracked by the existing inner current controller.

%

Voltage support is activated when \(V_{\mathrm{pu}}<0.85\) for 2 ms and released when \(V_{\mathrm{pu}}>0.92\) for 20 ms. The support current reference is limited to 1.0 pu, below the 1.25 pu device limit. During support and recovery, the battery converter regulates the dc link, while MC or loss of operating permission overrides the supervisor and forces its current references to zero. After MC reset, the references return to the native controller at 2 pu/s.

Each UPS battery is limited to 1 MW continuous discharge, 20 ms cumulative operation above 1 MW, 2 ms continuous operation above 1.5 MW, and a 600 MJ discharge-energy budget. These limits define the battery operating envelope used in this study. The 1050 MVA UPS aggregate and the 300 Mvar STATCOM are therefore compared as different support resources rather than equal-rated devices.

\subsection{Study Design and Performance Measures}


To evaluate the individual and combined effects of STATCOM support and UPS coordination, four configurations are tested. Configurations A/B use the original UPS control without/with STATCOM, while C/D use UPS coordination without/with STATCOM. Three-phase ground faults are applied at the midpoint of line 7--8 (near) or 4--5 (remote) at \(t_f=6.50\) s and cleared after 100, 150, or 200 ms. Simulations run to 11 s with a 10 \(\mu\)s EMT step and 100 \(\mu\)s output interval. Deep-sag A cases are used only to calibrate fault severity.


Changes in rectifier MC and auxiliary-load retention are bracketed by adjacent tested fault resistances. These intervals represent the tested transition ranges. 
To provide a common measure of fault severity, both endpoints of each MC transition interval are mapped to the residual PCC voltages under the baseline configuration (A) at the corresponding fault resistances.
The voltage is calculated as the time-averaged quadratic mean of the three phase RMS voltages over \(6.54\leq t<6.59\) s. 

Facility power is measured at the AIDC feeder, while UPS and STATCOM reactive powers are measured at the rectifier ac terminal and STATCOM transformer high-voltage side, respectively. AC powers use a 16.7 ms trailing average. The prefault power \(P_0\) is averaged over \(6.30\leq t<6.50\) s. Minimum grid draw and MC duration are evaluated over \(6.50\)--11.00 s, and the sustained recovery time is defined as
\begin{equation}
\Delta t_{90}=t_{90}-t_f,
\end{equation}
where \(t_{90}\) is the first time the facility power reaches \(0.9P_0\) and remains above it through 11 s. Battery discharge energy is defined as
\begin{equation}
E_{\mathrm{bat,+}}=
\int_{t_f}^{11\,\mathrm{s}}
\max\!\left(P_{\mathrm{bat,DCbus}}(t),0\right)\,dt .
\label{eq:batteryenergy}
\end{equation}

Additional electrical checks require the dc-link voltage to remain within 1.08--1.32 kV and the averaged internal-load power to remain above 90\% of its prefault value. 

\begin{figure*}[!t]
 \centering
 \includegraphics[width=\textwidth]{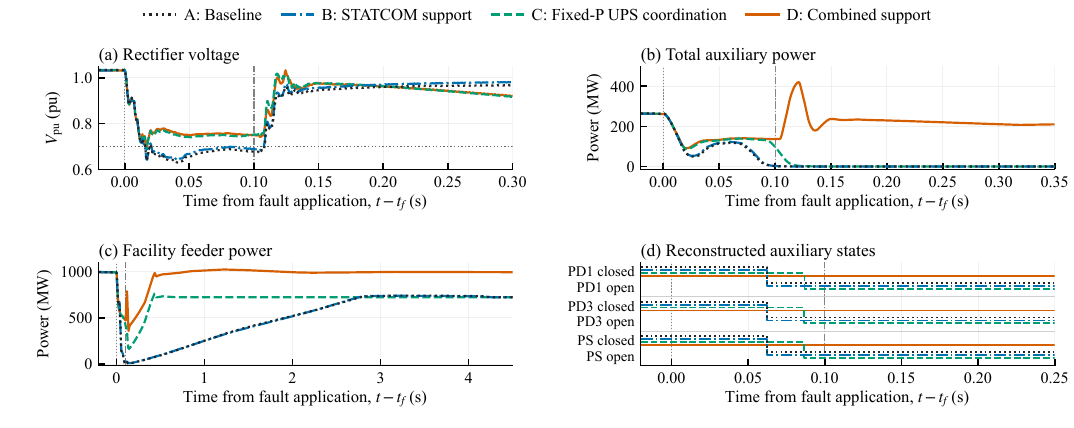}
 \caption{Responses of configurations A–D, near $R_f = 2.315 \Omega$, 100 ms: (a) rectifier voltage; (b) auxiliary power; (c) facility grid draw; (d) reconstructed auxiliary state. Both C and D avoid MC; only D retains all auxiliaries. Vertical lines mark fault commands.}
 \label{fig:auxiliary}
  \vspace{-2ex}
\end{figure*}

\section{Simulation Results and Discussion}
%
%

\subsection{MC Avoidance Under Different Fault Conditions}

For 100 ms three-phase ground faults, the near case with \(R_f=3.03~\Omega\) and the remote case with \(R_f=1.87~\Omega\) produce similar uncompensated PCC voltage voltage intervals of about 0.734 pu, where \(R_f\) denotes the fault resistance. However, the 300 Mvar STATCOM avoids MC only in the near case. In the remote case, the local rectifier voltage remains below 0.7 pu for 13.8 ms. Among the tested STATCOM ratings of 0, 200, 300, and 400 Mvar, 200 and 400 Mvar are the lowest ratings that avoid MC for the near and remote faults, respectively.

As shown in Fig.~\ref{fig:boundaries}, UPS coordination extends the tested MC-avoidance range to deeper near faults than STATCOM support alone. At \(R_f=2.42~\Omega\), both coordinated configurations avoid MC despite brief excursions of the local rectifier voltage below 0.7 pu, because the subthreshold durations are only 0.3 and 0.5 ms, well below the 10 ms pickup delay. The coordinated cases also avoid MC for the remote fault with \(R_f=1.87~\Omega\) (Table~\ref{tab:cases}). These results show that MC is governed by both the local voltage magnitude and its duration. The improvement with UPS coordination results from the combined effects of reduced active-power intake, capacitive reactive-current support, and battery buffering.

%
%

\subsection{Auxiliary-Load Retention and Facility Recovery}

Auxiliary-load protection introduces an additional constraint on facility recovery. For the 100 ms fault with \(R_f=2.315~\Omega\), both coordinated configurations avoid MC, but only combined support retains all three auxiliary branches (Fig.~\ref{fig:auxiliary}). Consequently, late facility demand reaches 993.39 MW with combined support, compared with 722.65 MW with UPS coordination alone. The approximately 263 MW difference is mainly associated with the retained auxiliary loads.

The milder 100 ms fault with \(R_f=3.03~\Omega\) shows a different tradeoff. 
STATCOM support alone maintains a minimum facility grid power intake of 696.08 MW, with only 7.159 MJ of battery discharge.
UPS coordination reduces the minimum demand to about 400 MW during the disturbance, but retains the auxiliary loads and restores the facility demand above 90\% of its prefault value within 0.394 s. In contrast, the original-control cases remain below this level at 11 s because the auxiliary trips remain latched. Thus, short-term demand retention and postfault recovery can favor different support strategies.

%
%
%
%
\begin{figure*}[!t]
 \centering
 \includegraphics[width=\textwidth]{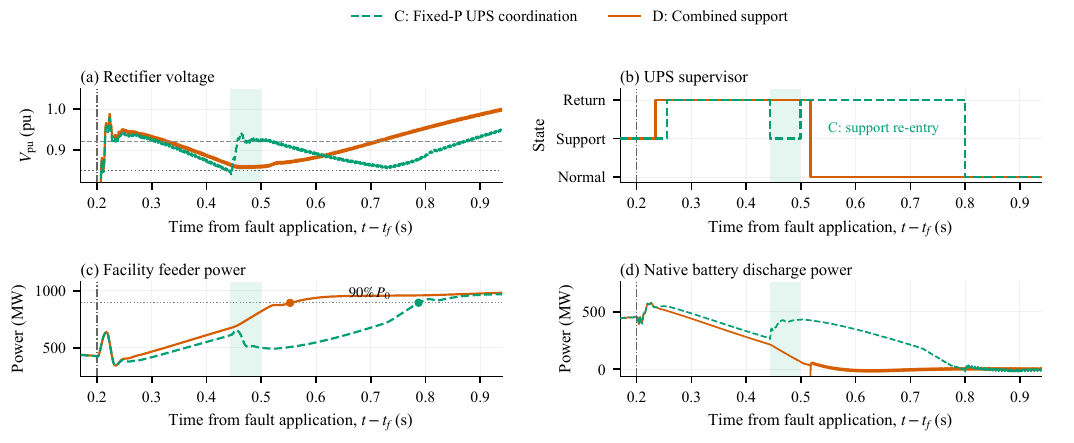}
 \caption{Recovery after the near 3.03 $\Omega$, 200 ms fault: rectifier voltage, supervisor state, facility grid draw, and battery discharge power. Shading marks C's support re-entry; voltage thresholds are 0.85/0.92 pu. Dots mark sustained 90\% recovery. Both configurations avoid MC and retain auxiliaries.}
 \label{fig:duration}
 \vspace{-2ex}
\end{figure*}

\subsection{Fault Duration and Postfault Recovery Dynamics}

For \(R_f=3.03~\Omega\), STATCOM support avoids MC for 100 and 150 ms faults, but a 200 ms fault produces 75.7 ms of MC (Table~\ref{tab:cases}). The cessation is triggered by the local undervoltage duration, while STATCOM protection remains inactive.

At 200 ms, both coordinated configurations avoid MC and retain the auxiliary loads, but their recovery differs (Fig.~\ref{fig:duration}). During recovery, the minimum local voltage is about 0.856 pu with combined support and 0.841 pu with UPS coordination alone. The latter falls below the 0.85 pu support threshold long enough to trigger a second support interval, which again reduces active intake and prolongs battery discharge.

As a result, combined support reduces full-event battery discharge from 293.549 to 193.705 MJ, a 34.01\% reduction, and advances sustained 90\% recovery from 0.7873 to 0.5528 s after fault application. These benefits arise from improved postfault recovery rather than MC avoidance, since neither coordinated case enters MC.

For the deeper \(R_f=1~\Omega\) fault, all four configurations enter MC, and the STATCOM-equipped cases exhibit longer MC durations. Thus, the benefit of voltage support depends on fault severity and recovery dynamics rather than increasing monotonically with added support.

%
%
%
\subsection{Performance Limits Under Deep Voltage Sags}

Two additional 100 ms near faults are used to examine the limits of combined support under unchanged control settings. Their uncompensated PCC voltage voltage intervals are approximately 0.5 and 0.3 pu, corresponding to \(R_f=1.42808\) and \(0.73849~\Omega\), respectively. With combined support, the PCC voltage intervals increase only slightly to 0.50839 and 0.30643 pu. MC still occurs for 149.3 and 394.7 ms, while the minimum facility demand falls to 3.53 and 1.26 MW.

Although the STATCOM current remains near 0.97 pu, its reactive-power output decreases to 147.8 and 89.5 Mvar as the terminal voltage falls. Battery support maintains the dc link, and the minimum internal IT load remains at 96.74\% and 95.85\% of its prefault value. However, all auxiliary branches trip and latch, and neither case recovers to 90\% of prefault facility demand by 11 s. These results show that combined support cannot prevent large grid-demand reductions under sufficiently deep voltage sags.

\section{Conclusion}

STATCOM and UPS-based voltage support improve AIDC responses to fault-induced voltage sags through different mechanisms. STATCOM better preserves short-term grid power intake with limited battery use under milder faults, whereas UPS coordination extends MC avoidance to deeper tested faults by reducing active intake, providing reactive support, and using battery buffering. Combined support further improves auxiliary-load retention and postfault recovery, including a 34.01\% reduction in battery discharge for the 200 ms case. However, sufficiently deep voltage sags still trigger MC and prevent full facility-demand recovery despite continued internal IT supply. These results show that AIDC voltage-support strategies should be evaluated not only by MC avoidance, but also by grid power draw, auxiliary-load retention, recovery dynamics, and battery requirements.

\bibliographystyle{IEEEtran}
\bibliography{references}
\end{document}